\documentclass[twocolumn]{aastex631}

\usepackage{CJK}
\usepackage{booktabs}
\usepackage{enumitem}

\begin{document}
\begin{CJK*}{UTF8}{gbsn}

\title{Disk dissipation, giant planet formation and star-formation-rate fluctuations in the past three-million-year history of Gould's Belt}

\correspondingauthor{Jinhua He(何金华)}
\email{jinhuahe@ynao.ac.cn}

\author[0000-0002-8898-3894]{Mingchao Liu (刘明超)}
\affiliation{Yunnan Observatories, Chinese Academy of Sciences, 396 Yangfangwang, Guandu District, Kunming, 650216, Yunnan, P. R. China}
\affiliation{Chinese Academy of Sciences South America Center for Astronomy, National Astronomical Observatories, Chinese Academy of Sciences, Beijing, 100101, P. R. China}
\affiliation{Departamento de Astronom\'{i}a, Universidad de Chile, Casilla 36-D, Santiago, Chile}
\affiliation{University of the Chinese Academy of Sciences, Yuquan Road 19, Shijingshan Block, Beijing, 100049, P. R. China}

\author[0000-0002-3938-4393]{Jinhua He (何金华)}
\affiliation{Yunnan Observatories, Chinese Academy of Sciences, 396 Yangfangwang, Guandu District, Kunming, 650216, Yunnan, P. R. China}
\affiliation{Chinese Academy of Sciences South America Center for Astronomy, National Astronomical Observatories, Chinese Academy of Sciences, Beijing, 100101, P. R. China}
\affiliation{Departamento de Astronom\'{i}a, Universidad de Chile, Casilla 36-D, Santiago, Chile}

\author[0000-0002-9600-1846]{Jixing Ge (葛继兴)}
\affiliation{Chinese Academy of Sciences South America Center for Astronomy, National Astronomical Observatories, Chinese Academy of Sciences, Beijing, 100101, P. R. China}

\author[0000-0002-5286-2564]{Tie Liu (刘铁)}
\affiliation{Shanghai Astronomical Observatory, Chinese Academy of Sciences, 80 Nandan Road, Shanghai, 200030, People's Republic of China}

\author[0000-0002-6581-3307]{Yuping Tang (唐雨平)}
\affiliation{Chinese Academy of Sciences South America Center for Astronomy, National Astronomical Observatories, Chinese Academy of Sciences, Beijing, 100101, P. R. China}

\author[0000-0002-2783-2080]{Xuzhi Li (李旭志)}
\affiliation{Deep Space Exploration Laboratory/Department of Astronomy, University of Science and Technology of China, Hefei, 230026, China.}

\affiliation{School of Astronomy and Space Sciences, University of Science and Technology of China, Hefei, 230026, China}
 
\begin{abstract}

Although episodic star formation (SF) had been suggested for nearby SF regions, a panoramic view to the latest episodic SF history in the solar neighborhood is still missing. 
By uniformly constraining the slope $\alpha$ of infrared spectral energy distributions (SEDs) of young stellar objects (YSOs) in the 13 largest Gould's Belt (GB) protoclusters  surveyed by Spitzer Space Telescope, we have constructed a cluster-averaged histogram of $\alpha$ representing YSO evolution lifetime as a function of the $\alpha$ value. 
Complementary to the traditional SED classification scheme (\textsc{0,i,f,ii,iii}) that is based on different $\alpha$ values, a staging scheme (A,B,C,D,E) of SED evolution is advised on the basis of the $\alpha$ statistical features that can be better matched to the physical stages of disk dissipation and giant planet formation. 
This has also allowed us to unravel the fluctuations of star formation rate (SFR) in the past three-million-year (3\,Myr) history of these GB protoclusters.
Diverse evolutionary patterns such as single peak, double peaks and on-going acceleration of SFR are revealed.
The SFR fluctuations are between $20\%\sim60\%$ ($\sim40\%$ on average) and no dependence on the average SFR or the number of SFR episodes is found. However, spatially close protoclusters tend to share similar SFR fluctuation trends, indicating that the driving force of the fluctuations should be at size scales beyond the typical cluster sizes of several parsec.

\end{abstract}



\section{Introduction} \label{sec:intro}

Most stars formed in clusters \citep{Lada2003,Porras2003}, but it is still unclear how star clusters form in molecular clouds. 
Recently there has been increasing evidence that star formation in clusters is episodic.
For example, multiple episodes of star formation or apparent acceleration of star formation have been revealed in the Hertzsprung-Russell (HR) diagram of some nearby protoclusters such as Orion Nebula Cluster, Taurus-Auriga, Lupus, Chamaeleon\,\textsc{i}, $\rho$\,Ophiuchi, IC\,348, and NGC\,2264  \citep{Palla2000,Huff2006,Spezzi2008,Beccari2017}. 
The infrared (IR) photometry by Spitzer Space Telescope has enabled the discovery of one or two peaks in the histograms of IR SED slope $\alpha$ in nearby protoclusters such as Chamaeleon\,\textsc{i} \citep{Luhman2008},  Auriga-California Molecular Cloud \citep[AurigaCMC;][]{Broekhoven2014}, Cepheus \citep{Kirk2009}, Ophiucus \citep{Alves2012}, Perseus \citep{Azimlu2015} and Taurus \citep{Luhman2010},  which indicates diverse star formation (SF) histories in these regions. 
\citet{Hsieh2013} even briefly compared the $\alpha$ histograms of several nearby clouds. 
However, these $\alpha$ peaks are usually either not discussed in depth or simply attributed to SF episodes. The important fact that disk lifetime can be very different at different $\alpha$ values is usually neglected.
Improved astrometric measurements provided by the GAIA satellite have revealed multiple episodes of SF also in older populations of some nearby SF regions such as Aquila/Serpens \citep{Herczeg2019}, Lupus \citep{Galli2020}, and the newly identified Orion C \citep{Kounkel2018}. 
It is thus timely to exploit the power of low mass stars from large surveys \citep{Megeath2022} to obtain a panoramic view to the episodic SF history in the earliest stage in the nearby protocluster forming regions.

The infrared SEDs of YSOs were classified using their slopes $\alpha$ by \citet{Lada1987,Adams1987} and later improved by \citet{Andre1993,Greene1994} into Class\,0, \textsc{i}, \textsc{f}, \textsc{ii}, and \textsc{iii}, which are roughly linked to the evolutionary sequence of YSOs. The $\alpha$ slope is a good tracer of the evolutionary time scales of circumstellar disks. \citet{Wilking1989}  tried to assign Kelvin-Helmholtz timescales for the different SED classes, particularly the 2\,Myr lifetime for Class\,\textsc{ii} sources. The life times of the SED classes were further refined by \citet{Evans2009} using the \textit{Spitzer c2d} survey (PI: N.~J. Evans) of nearby protostars. 

However, our unpublished preliminary analysis of the star number counts of the different SED classes in the GB protoclusters has shown limitations in using the SED classification scheme to trace SF history. 
For example, the fractions of Class\,\textsc{f} YSOs seem to be correlated with that of Class\,\textsc{0+i} ones among the GB protoclusters, which means the Class\,\textsc{f} is perhaps not a distinctive evolutionary stage of YSOs than Class\,\textsc{0+i}.
This point will be also confirmed by the correlation analysis in Appendix~\ref{app:alpha_time_error_bar}.
In this work, such limitations will be overcome by directly using the statistics of $\alpha$, instead of the traditional SED classification, in the analysis of YSOs evolution.

In this paper, we will introduce the GB protoclusters from literature and the reprocessing of their IR SEDs in Sect.~\ref{sec:data}, analyze the $\alpha$ histograms, convert them into disk-evolutionary-age histograms, discuss the obtained disk-age histograms and average $\alpha$ histogram in Sects.~\ref{subsec:age_old_new} to \ref{subsec:age_histograms}, define the SED evolution stages and match them to disk dissipation and giant planet formation stages in Sect.~\ref{subsec:disk_dissipation_planet}, and finally summarize the main findings in Sect.~\ref{sec:summary}.

\section{Star samples and data processing} \label{sec:data}

\subsection{Star samples and IR data} \label{subbsec:data_IR}

In order to understand the protoclusters in Gould's Belt as a whole, we collect published IR photometry between 2 and 24\,$\mu$m of all GB protoclusters in the catalog papers, \citet{Rebull2010,Megeath2012} of \textit{Spitzer Taurus} and \textit{Orion} surveys and  \citet{Dunham2015} of the \textit{Spitzer c2d} (PI: N.~J. Evans) and \textit{Spitzer GB} (PI: L.~E. Allen) surveys. 
These works have utilized color and magnitude criteria and, in some cases, spectroscopy to identify YSOs, excluded most background galaxies, outflow knots and part of field giant stars that also show IR excess, and matched the sources to 2MASS catalogue \citep{Skrutskie2006}. Several protoclusters in these papers are composed of more than one spatially well separated parts. We treat each part as a single protocluster and use their orientations (e.g., E for East, W for West, NE for Northeast) to differentiate them. 

However, the contamination of giant stars to the Class\,\textsc{iii} YSOs is still an issue in these catalogs. 
We thus use the GAIA distances to identify and remove potential giant star contaminants as much as possible. For this purpose, all the above Spitzer YSOs are checked in the CDS X-Match Service \footnote[1]{http://cdsxmatch.u-strasbg.fr/\#tab=xmatch\&}\citep{Pineau2020}
to search for their GAIA counterparts in the Early Data Release 3 (EDR3) catalog \citep{Lindegren2021} . 
A matching radius of $2''$ is used. We have found 4181 one-to-one matches among the total of 6443 candidate member stars. 
For 299 multiple-matching cases, we select the closest GAIA source for the YSOs. 
Usually, mainly Class\,\textsc{ii} or \textsc{iii} YSOs are found to have GAIA counterparts. For the remaining 1963 candidate stars that have no GAIA counterparts, they are very likely embedded sources in the dense clouds so that we do not have accurate distance for them; we simply adopt them as member stars.

We adopt the following conservative criterion to identify nonmembers of each protocluster. 
First we observe the GAIA distance distribution of each protocluster by eyes and manually determine a reasonable distance range for member stars. 
In most cases, member stars distribute in a single continuous range of distances. 
However, in the few protoclusters near to Galactic plane (Aquila-Main, Serpens-Main and Serpens-NE), the candidate member stars gather around two distinct distance ranges. In these cases, we select the closer distance range for the protocluster, because the farther one is very likely dominated by background giant stars (contaminants). 
The adopted distance ranges are listed in Appendix Table~\ref{tab:dist_range}. 
If a star's distance, including the corresponding distance uncertainties, lies entirely outside this distance range, then it is identified as a nonmember and excluded from our analysis. As a result, we have excluded $3\sim354$ contaminants from each of the 13 largest (with $\ge50$ final member stars) GB protoclusters (the detailed numbers of foreground and background contaminants are also give in Appendix Table~\ref{tab:dist_range}).
The other smaller GB protoclusters will not be considered further in this work.
We have noticed that the adopted distance ranges are usually much larger than the transverse sizes (several pc) of the corresponding protoclusters, which we temporarily attribute to the large uncertainties of GAIA distances. 
We have examined the GAIA proper motions of the selected member stars and confirmed that the majority of them always concentrate well in a single small region.

The catalog paper of \citet{Dunham2015} also provided the extinction corrected SED slope $\alpha$ for the cluster member stars. 
However, as we will discuss in next section, the derived $\alpha$ histograms showed bewilderingly diverse peak positions, which is hard to interpret from a theoretical point of view. 
Thus, we decide to reconstruct the SEDs starting from the IR fluxes in the three catalog papers, using the extinction law of \citet{Weingartner2001}, and compute the SED slope $\alpha$ in a uniform manner for stars from all the three papers.

\subsection{Extinction correction} \label{subsec:data_extinction}

The catalog papers of the GB protoclusters have estimated foreground extinction for each member star. 
In each SF region, the extinction was mainly determined for sources of SED Class\,\textsc{ii} and \textsc{iii} and then a uniform average extinction from these sources was adopted for embedded sources of Class\,\textsc{0, i} and \textsc{f}. 
Therefore, the extinction in the literature mainly reflects the extinction of foreground dust. 
We adopt the extinction estimates $A_V$ or $A_K$ from the catalog papers, but use the extinction curve of \citet[][; the extinction curve for $R_V=5.5$ for molecular clouds and Case B grain size distribution are obtained from a website\footnote{The model extinction curve is from:\\ \url{https://www.astro.princeton.edu/~draine/dust/dustmix.html}} ]{Weingartner2001} to convert $A_V$ or $A_K$ to the extinction in the other IR bands. 
Therefore, we need not to make the correction for scattering separately, as did by  \citet{Dunham2015}. 
We caution that the local extinction of embedded sources still remains after this step.

\subsection{Saturation issues} \label{subsec:data_saturation}

Most of the GB protoclusters have the Ks band and 24\,$\mu$m photometry available for almost all their member stars, with the exception of Orion A, B-M and B-S due to a saturation issue in the \textit{Spitzer}/MIPS maps. We have tried to recover the saturated 24\,$\mu$m fluxes from their empirical correlations with the fluxes at shorter wavelengths among the unsaturated sources for 30\%, 21\% and 43\% of member stars of the above three protoclusters respectively (see the details in Appendix~\ref{sec:24umflux}). A  portion of member stars (10\%, 3\% and 19\%, respectively) still have to be abandoned because photometry is absent for all wavelengths $\lambda\ge5.8\,\mu$m. In addition, we drop a small number of member stars whose 24\,$\mu$m flux is missing in several other protoclusters; the numbers of dropped stars are 9 (Ophiuchus), 2 (Serpens-Main), 8 (Taurus) and 5 (Perseus-W), which does not impact the statistics.

\subsection{Fit the SED slope} \label{sec:data_fit_alpha}

The slope $\alpha$ of an IR SED is derived by fitting a straight line to it in the $\log{(\lambda F_\lambda)}$ vs $\log{\lambda}$ plot. 
We choose to fit the SED from the 2MASS Ks band (2.2\,$\mu$m) to Spitzer MIPS 24\,$\mu$m band because the former traces the hot dust in the inner disk and the latter traces the cold dust in the outer disk (and envelop in earliest stages) so that the derived $\alpha$ will be a good tracer of the evolution of all circumstellar matter. 
For 528 stars that have the K-band flux missing (nearly 80\% of them are Class~\textsc{0, i, f}, likely due to extinction), their SEDs start from the IRAC 3.6$\mu$m band. This does not impact the $\alpha$ values much. 
In addition, following this traditional definition of $\alpha$ offers us the opportunity to use the classical lifetime of 2\,Myr of Class\,\textsc{ii} YSOs \citep{Wilking1989,Evans2009} to do an absolute calibration to the timescale in the $\alpha$-age conversion relation (see 
details in Sect.~\ref{subsec:alpha_to_age}). 

\section{Results and discussions} \label{sec:age}

\subsection{Two versions of SED-slope histograms} \label{subsec:age_old_new}

\begin{figure}[ht!]
\epsscale{0.9} 
\plotone{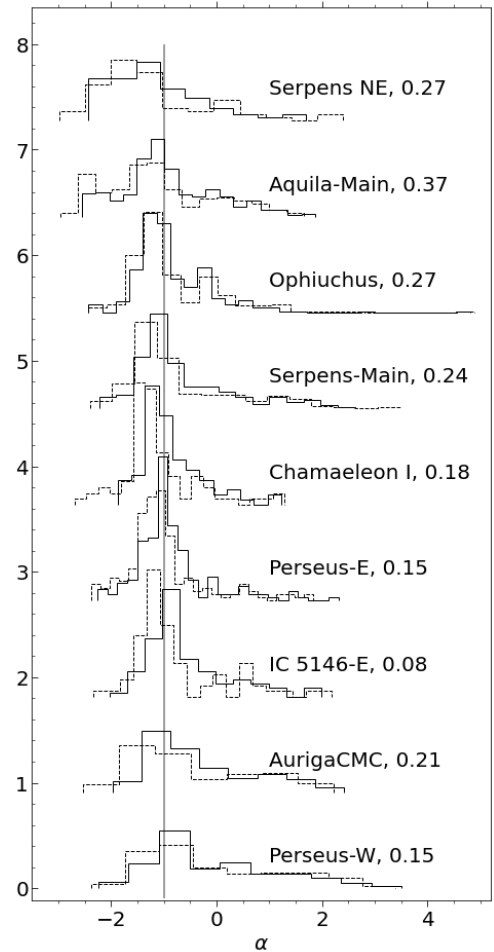}
\caption{Compare the histograms of $\alpha$ values from literature \citep[][; dashed lines]{Dunham2015} and from this work (full lines) for the nine protoclusters with $\ge50$ member stars in that paper. All histograms are normalized to an area of unity and vertically shifted for clarity. The number after each cluster name is the average amount of decrease of the $\alpha$ values of all member stars of the corresponding protocluster due to our extinction correction. The vertical gray full line marks the average $\alpha$ peak position of our version of histograms to facilitate comparison.
\label{fig:comp_alpha_histo}}
\end{figure}

We initially used the extinction corrected $\alpha$ values from \citet{Dunham2015} to make the histograms. However, the peak positions of the histograms turned out to be significantly different among the protoclusters (the dashed histograms in Fig.~\ref{fig:comp_alpha_histo}). 
We stress that, because the histogram bins are very broad for smaller clusters, the shift of the $\alpha$ peak by one bin is already very large.
This is hard to understand from a theoretical point of view because, if the evolution of SED is dominated by the common disk physics, as expected, the $\alpha$-histograms of all protoclusters should peak around the same $\alpha$ value that corresponds to the slowest evolutionary stage. Only extremely sharp star formation rate (SFR) fluctuations have the potential to significantly shift the peak position, which means that the largest peak shift should appear in the narrowest $\alpha$ histogram. However, this is not the case in Fig.~\ref{fig:comp_alpha_histo}. For example, Serpens\,NE has the largest peak shift  with respect to the others (among the dashed line histograms) but has a very broad $\alpha$ peak.

In order to resolve this problem, we start from the IR fluxes in the literature and redo the extinction correction and SED fitting in a uniform way for all protoclusters, as described in Sect.~\ref{subsec:data_extinction}. 
We compare the derived $\alpha$-histograms with that made from the literature $\alpha$ values in Fig.~\ref{fig:comp_alpha_histo} for the nine protoclusters that have $\ge50$ member stars in \citet{Dunham2015}. 
The effects of our extinction correction to these stars are the decreases of $\alpha$ values by $0.08\sim0.37$ on average (see the numbers after the cluster names in the figure).
The histograms of the literature $\alpha$ values (dashed lines)  generally peak at smaller $\alpha$ values than ours (full lines) and these peak shifts are roughly comparable to or even larger than the average effects of the extinction correction itself for many of these protoclusters, which reflects the over-correction of the dust-scattering extinction by \citet{Dunham2015}.


\subsection{All the SED-slope histograms} \label{subsec:age_alpha_histograms}

\begin{figure*}[!htb]
    \centering
    \includegraphics[width=0.8\textwidth]{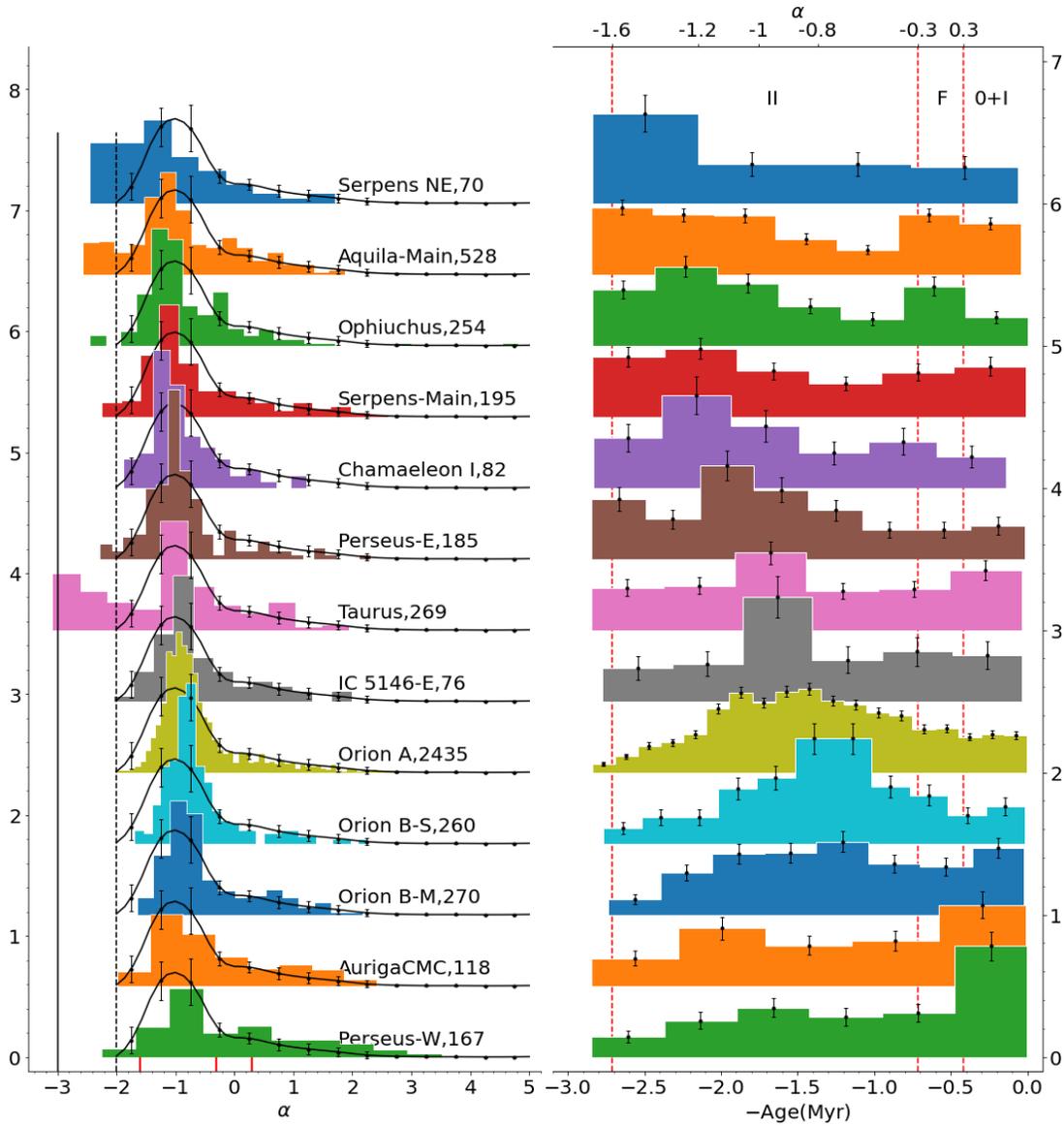} 
    \caption{Histograms of spectral energy distribution (SED) slope $\alpha$ {\bf (Left)} and disk age {\bf (Right)} of YSOs of the 13 largest protoclusters in Gould's Belt as surveyed by Spitzer Space Telescope. The number of member stars is shown behind each object name. The black curve in the left panel is a smoothed version of the average histogram of the 13 protoclusters. The averaging was only performed to the right side of  the vertical black dashed line, i.e. within $-2\le\alpha\le5$, and the common bin width for averaging is $\Delta\alpha=0.5$. 
    All histograms are normalized to an area of unity in the considered value range and vertically shifted for clarity.
    The vertical black full line marks the theoretical limit of $\alpha=-3$ for main sequence stars much hotter than 3000\,K. 
    The disk-evolution-time histograms in the right panel are converted from the $\alpha$ histograms according to Sect.~\ref{subsec:alpha_to_age}. 
    Negative time means in the past and 0\,Myr is now. 
    The error bars are the dispersion among the 13 protoclusters for the black curve in the left panel and the square root of star numbers in each time bin in the right panel. 
    Both the short vertical red lines on the $\alpha$-axis in the left panel and the long vertical red dashed lines in the right panel mark the border lines of traditional classification of protostar SED types: \textsc{0+i}, \textsc{f}, \textsc{ii} and \textsc{iii}. The bin width of each histogram is optimized with the `Freedman-Diaconis' rule \citep{Freedman1981OnTH}.}
    \label{fig:alpha_histo}
\end{figure*}

Our version of $\alpha$-histograms of all the 13 GB protoclusters, including those in Orion and Taurus SF regions, are shown together in the left panel of Fig.~\ref{fig:alpha_histo}. 
The most salient feature in the figure is that all the protoclusters show the major peak around $\alpha\sim-1$, indicating that these histograms mainly depend on the disk lifetime in different evolutionary stages traced by $\alpha$.
However, individual protoclusters also show secondary diversities in the histograms. 
For example, the major-peak positions can vary by $\Delta\alpha\sim0.2$ from cluster to cluster and several protoclusters show a sharp peak (e.g., Chamaeleon\,\textsc{i}, IC\,5146-E, Orion B-S and Perseus-E), hinting on bursty SF events. 
A few protoclusters also possess older Class\,\textsc{iii} member stars with $\alpha<-2$ (e.g., Serpens NE, Aquila-Main, Ophiuchus and Taurus) which means that their recent star formation might be initiated earlier than the other protoclusters.

\subsection{Conversion between SED slope and disk age} \label{subsec:alpha_to_age}

The good agreement of the $\alpha$ histrograms among all the 13 GB protoclusters allows us to derive the disk evolution lifetime in  the range of each bin of the histograms by averaging the histograms (after binned to the same $\alpha$ grid with bin width $\Delta\alpha=0.5$) among the clusters to suppress the minor diversities. 
The so derived average $\alpha$ histogram, after a spline interpolation to a finer $\alpha$ grid to facilitate later analysis, is shown as the black curves in the left half of Fig.~\ref{fig:alpha_histo} and the values of the average $\alpha$ histogram are listed in Appendix Table~\ref{tab:alpha_time} to facilitate its reuse in future works. 
The difference of individual histograms with respect to the average can be interpreted as due to the temporal variation of SFR of each individual protocluster. 
The peaking of $\alpha\sim-1$ was found by other authors in individual protoclusters such as AurigaCMC \citep{Broekhoven2014}, Cepheus \citep{Kirk2009}, Chamaeleon\,\textsc{i} \citep{Luhman2008}, Ophiuchus \citep{Alves2012}, Perseus \citep{Azimlu2015}, and Taurus \citep{Luhman2010}, but little attention was paid to test its link to disk evolution lifetime. 

However, there are several weaknesses in the above treatment. 
First, if all the protoclusters somehow share the same temporal variation trend in $\alpha$ during their formation, this common trend will enter the averaged histogram and alter the derived disk evolution lifetimes. 
Second, the number of protoclusters is still small so that uncertainties are still large in the average $\alpha$ histogram. 
Thirdly, there exist SED diversities either due to different disk evolution paths \citep{Strom1989,Lada2006,Cieza2007} or due to random disk inclination angles \citep{Whitney2003,Crapsi2008}.
Finally, the number of the youngest member stars (with $\alpha>1$) could be underestimated due to large extinction to their locations in the dense clouds.

Despite these weaknesses, it is still instructive to integrate the distribution of disk-evolution lifetimes (represented by the $\alpha$-bin heights) to transform the $\alpha$ histograms into that of the evolution age of an ``average disk''. 
To integrate the average histogram from $\alpha=5$ to any $\alpha$ value between $-2$ and $5$, we first fit the histogram with a second order spline function using the \textit{splrep} and \textit{splint} functions of the \textit{scipy.interpolate} B-spline package (Note: A cubic spline will produce an unrealistic oscillation). The spline fit is denoted as $P(\alpha)$ and normalized into a probability density function as $\int_{-2}^5{P(\alpha)d\alpha}=1$. Then, the disk life time in any unit $\alpha$ range of $d\alpha$ is $dt=aP(\alpha)d\alpha$, where $a$ is a constant scaling factor to be calibrated. The conversion relation from $\alpha$ to disk evolution time $T$ (or negative disk age) can be numerically derived as 
\begin{equation}
    \label{eq:alpha-t}
    T(\alpha)=\int_0^Tdt=a\int_5^\alpha P(\alpha)d\alpha
\end{equation}
for $-2\le\alpha\le5$. The constant $a\approx2.86$\,Myr is obtained by adopting the traditional assumption of a lifetime of 2\,Myr for Class\,\textsc{ii} objects \citep[$-1.6\le\alpha<-0.3$;][]{Wilking1989,Evans2009}, i.e., $T(-0.3)-T(-1.6)=2.0$. 

A graphic presentation of Eq.~\ref{eq:alpha-t} is given in Fig.~\ref{fig:alpha_to_age} and the data of this curve are given in Appendix Table~\ref{tab:alpha_time} for reuse in future works. 
The error bars are numerically evaluated by a Monte Carlo procedure in Appendix~\ref{app:alpha_time_error_bar}. 
We can see that, for an average disk, the $\alpha$ value first decreases quickly with time in the early stages when $\alpha\ge2.5$. 
Then, the decrease of $\alpha$ gradually slows down, reaches a slowest point around $\alpha=-1$ and speeds up again towards $\alpha=-2$. 
These trends have fundamental indications to the dissipation rate of circumstellar dust at a population level, which we will discuss in detail in Sect.~\ref{subsec:disk_dissipation_planet}. We stress that this conversion relation is less meaningful for individual stars than for star populations because of SED diversities, as we will discuss later.

\begin{figure}[ht!]
\epsscale{0.8} 
\plotone{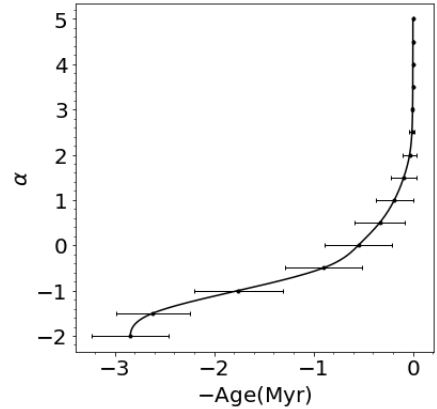}
\caption{The conversion relation from SED slope $\alpha$ to disk evolutionary age for YSOs, as expressed by Eq.~\ref{eq:alpha-t}. The error bars are numerically obtained as in Appendix~\ref{app:alpha_time_error_bar}. The data behind the plot is given in Appendix Table~\ref{tab:alpha_time}.
\label{fig:alpha_to_age}}
\end{figure}

Applying Eq.~\ref{eq:alpha-t} to every member star of all the GB protoclusters, we derive the histograms of disk evolutionary ages in the right half of Fig.~\ref{fig:alpha_histo}. The error bars of the disk-age histograms come from the square root of star number of each bin. We do not transfer the error bars in Fig.~\ref{fig:alpha_to_age} into the disk-age histograms, because these error bars reflect both the star-number counting error of each $\alpha$ histogram and the error due to the fluctuations of SFR,
while the purpose of the right half of Fig.~\ref{fig:alpha_histo} is merely to present the same SFR fluctuations through directly comparing the disk-age histograms. 
However, if these disk-age histograms would be compared with similar histograms obtained with other methods in future, the error bars in Fig.~\ref{fig:alpha_to_age} should be transferred to the disk-age histograms through another simple Monte Carlo approach to replace the current number counting error bars.

We comment that the derived disk-age histograms look very different from the original $\alpha$ histograms in the left half of Fig.~\ref{fig:alpha_histo}. This is because, on the one hand, the corresponding ranges of disk-age are distinct in different $\alpha$ bins and, on the other hand, the disk-age histograms only reflect the differences between the individual $\alpha$ histograms and the averaged one. 

We also remark that our results indicate apparent disk-age uncertainties better than 0.5\,Myr in the youngest stage of $<3$\,Myr, as can be seen through the error bars in Fig.~\ref{fig:alpha_to_age}. 
However, we have implicitly assumed that all YSOs have the same disk-evolution track traced by $\alpha$. If other factors such as 
the different stellar mass, disk mass, accretion history, metallicity, initial angular momentum, mass reservoir, etc. would be taken into account, the absolute disk-age uncertainties will be larger. The situation is similar to stellar ages constrained through the Hertzprung-Russell Diagram (HRD) for similarly young stars where uncertainties in the theoretical models make the stellar age less accurate than apparently determined on the HRD \citep{Soderblom2014}.
As we will discuss in a later subsection, some of these neglected factors are responsible for the observed SED diversities.


We caution again that the sample of the youngest and reddest member stars are usually embedded in dense cloud cores and are thus incomplete due to high local extinction. Thus, the real evolution of the SED slope during the youngest stages could be slower than we found here.

\subsection{Disk-age histograms} \label{subsec:age_histograms}

The derived disk-age histograms in the right half of Fig.~\ref{fig:alpha_histo} clearly demonstrate, for the first time, the acceleration and/or deceleration of SFR in the latest 3\,Myr history of each protocluster. 
Some protoclusters (Serpens NE, Serpens-Main, Aquila-Main, Ophiuchus and Chamaeleon\,\textsc{i}) are older, experienced their peak SF epoch more than 2\,Myr ago. 
Some younger protoclusters (Perseus-E, IC\,5146-E, Orion A, B-M, B-S and the major peak of Taurus) reached the maximum SFRs between $1\sim2$\,Myr ago. 
The two youngest protoclusters, Perseus-W and AurigaCMC, have their SFRs accelerating up to now. 
Some protoclusters seem to show multiple peaks in the disk-age histograms. 
A dip test of \citet{Hartigan1985}\footnote{We used the python code diptest:\\ \url{https://github.com/alimuldal/diptest}} is performed to the disk-ages of each protocluster. 
If the p-value is $<0.05$, the tested sample is multi-modal, otherwise if the p-value is $<0.1$, the tested sample possibly has multiple modes.
Aquila-Main ($p=0.0$), Ophiuchus ($p=0.004$) and Taurus ($p=0.008$) are found to be multi-modal and Serpens-Main ($p=0.09$) possibly multi-modal, indicating multiple SF episodes even within the short period of 3\,Myr. Combining these results with episodic SF events discovered on longer timescales in older Pre-Main Sequence (PMS) populations selected from other surveys than Spitzer \citep[e.g.,][]{Beccari2018,Herczeg2019} 
highlights a picture of multi-timescale episodic star formation.

An interesting feature in Fig.~\ref{fig:alpha_histo} is that spatially close protoclusters tend to share similar SFR flucturation trends. For example, Serpens\,NE, Aquila-Main, Serpens-Main and Ophiuchus protoclusters are close to each other on the sky plane  and all of them reached the SFR peak early (about 3\,Myr ago). 
The Orion A, B-S and B-M protoclusters are also close neighbors and their SFRs peak around the similar time of $\sim$1.5\,Myr ago. 
The AurigaCMC and Perseus-W protoclusters are also not far from each other and both of them are experiencing the acceleration of SF even up to today. 
The two protoclusters with the most prominent double peaks in the disk-age histograms, Aquila-Main and Ophiuchus, are also in adjacent regions. These features indicates that the fluctuations of SFR should be driven by agents at large size scales covering several adjacent protoclusters.

However, we also note that spatially close protoclusters do not necessarily always have similar SFR fluctuation trends. For example, Perseus\,E and W show the SFR peak at quite different times. This is also consistent with the diversity of SFR trends among sub-clusters within the same protoclusters, which we will discuss in our next paper. It hints on the complexity and diversity of cluster formation.

\begin{figure}[ht!]
\epsscale{1.2} 
\plotone{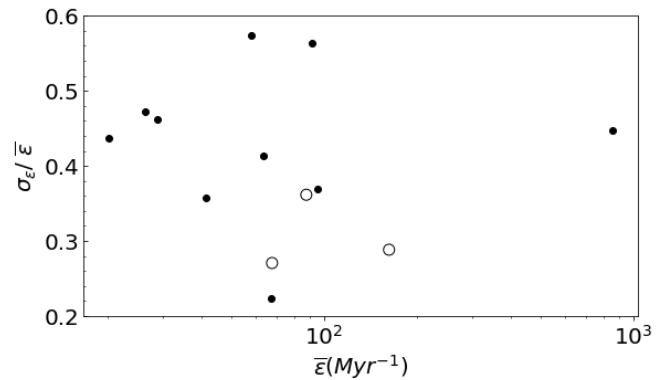}
\caption{The relationship between the average SFRs $\bar{\epsilon}$ (number of stars per Myr)  and its relative fluctuations (standard deviation divided by average)  $\sigma_\epsilon/\bar{\epsilon}$ among the 13 Gould's Belt protoclusters. The dots are clusters with a single peak in the disk-age histograms, while the circles are those with double peaks.
\label{fig:SFR_std}}
\end{figure}

In order to quantify the fluctuations of the SFRs ($\epsilon$), we compute the standard deviation of each disk-age histogram and show their ratios to the average height of the corresponding histograms in Fig.~\ref{fig:SFR_std}. 
Because the disk-age histograms are normalized to total areas of unity, the height of each bin is proportional to the level of SFR in each disk-age bin. 
Thus, the standard-deviation-to-average ratio of a disk-age histogram is also the ratio of the standard deviation ($\sigma_\epsilon$) of its SFRs to its average SFR ($\bar{\epsilon}$) or, in another word, its relative fluctuation of SFR ($\sigma_\epsilon/\bar{\epsilon}$). 
One can convert the SFRs into a unit of M$_\odot$\,yr$^{-1}$ by multiplying them with an average stellar mass of 0.5\,M$_\odot$.
It can be seen that the SFRs fluctuate strongly with amplitudes between $20\%$ to $60\%$ ($\sim40\%$ on average) and the fluctuations are not correlated with the average SFRs, nor with the number of peaks in the disk-age histograms. 
On the other hand, despite the large SFR fluctuations, all protoclusters maintain a non-negligible level of SF most of the time so that the SF should be a continuous activity within timescales of 3\,Myr, instead of neat discrete bursts. 

\subsection{A uniform statistical view of disk dissipation and giant planet formation} \label{subsec:disk_dissipation_planet}

In this section, we will first show that the shape of the average $\alpha$ histogram has not been impacted much by the sample selection biases. 
Then, we will introduce a new SED staging scheme that has the potential to neatly match the statistical features in the average $\alpha$ histogram to individual physical stages of disk dissipation. Finally, we will propose a scenario to naturally integrate the giant planet formation into the new scheme to play a key role in the disk dissipation.

\subsubsection{Sample biases} \label{subsubsec:Sample_bias}

For the first time, we have been able to split the effects of disk evolution timescales and fluctuations of SFR in the   $\alpha$ histograms of protoclusters at a population level. The common trends in the  $\alpha$ histograms in Fig.~\ref{fig:alpha_histo} or the $\alpha$-age conversion curve in Fig.~\ref{fig:alpha_to_age} shall cast new light on the dissipation processes of circumstellar dust around YSOs. 
However, before discussing that, we need to clarify how these trends have been distorted by selection effects. 
Beside the possible incompleteness of the youngest member stars (with $\alpha>0$) due to high (uncorrected) opacity in the dense molecular clouds, as mentioned in Sect.~\ref{subsec:data_extinction}, we will discuss another two biases below.

The first possible bias is that some YSOs might have been removed together with the background galaxies. Compared to typical YSOs, galaxies usually show redder colors at a given 24\,$\mu$m flux level \citep[e.g.,][]{Harvey2007}. Thus, the confusion with galaxies might have enhanced the incompleteness of the reddest YSOs of Class~0 and \textsc{i} to some extent. However, because the YSOs and galaxies are separated well on the multiple color-magnitude plots, this bias should be small.

The second bias is related to the removal of main sequence (MS) stars when selecting YSO candidates from the original \textit{Spitzer} maps. For example, when \citet{Harvey2007} selected YSOs in the Serpens regions, about 94\% of the Spitzer point sources with sufficient fluxes to make informative SEDs had been removed as MS stars. 
According to their approach \citep[described by][]{Evans2003}, all good fits with a theoretical photosphere model and the extinction law of \citet{Weingartner2001} are considered as MS stars. 
This means the removed MS stars should always have their de-reddened IR SEDs very similar to a blackbody function whose IR SED slope is theoretically $\sim-3$ and this bias mainly impacts this $\alpha$ range. 
However, the fact that the Taurus protocluster in our sample has shown a secondary $\alpha$ peak just around $\alpha=-3$ in the left half of Fig.~\ref{fig:alpha_histo} indicates that this bias is not necessarily very serious even in this $\alpha$ range. Therefore, because our discussions will be focused in the $\alpha\ge-2$ range, we need not worry about this bias too much.

\subsubsection{Stages of SED evolution and disk dissipation} \label{subsubsec:SED_staging}

\begin{figure}[ht!]
\epsscale{1.2} 
\plotone{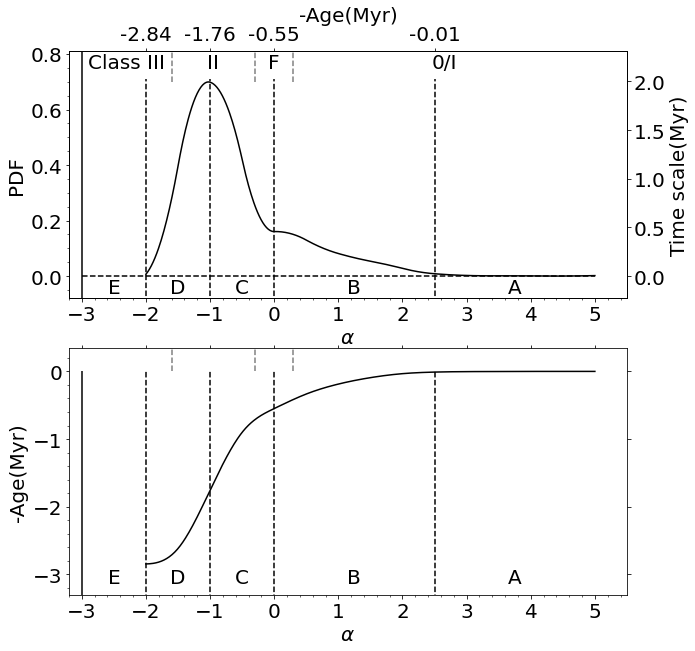}
\caption{SED evolutionary stages (A, B, C, D and E) defined on the basis of the statistical features of the average $\alpha$ histogram (\textit{top panel}, expressed in both probability density in the left Y-axis and disk evolution time scale per unit change of $\alpha$ in the right Y-axis) and the $\alpha$-age conversion curve (\textit{bottom panel}). The dashed border lines are at $\alpha=2.5$, 0.0, -1 and -2, or trace-back disk ages of -0.01, -0.55, -1.76 and -2.84\,Myr, respectively. The vertical full line marks the theoretical limit value $\alpha=-3$ of atmosphere (blackbody). The traditional SED classification borders at $\alpha=0.3$, -0.3 and -1.6 are also marked out for comparison. 
\label{fig:new_SED_classes}}
\end{figure}

It is thus reasonable to assume that the average $\alpha$ histogram is roughly free of observation biases and a physical interpretation is warranted. 
To facilitate discussions, we re-plot the smoothed average $\alpha$ histogram and the $\alpha$-age conversion curve together in Fig~\ref{fig:new_SED_classes}.
Because the $\alpha$ value is mainly determined by the K-band and/or 3.6\,$\mu$m emission from the hot inner disk (or from photosphere in the case of transition disk) and the 24\,$\mu$m emission from the outer cooler disk beyond $5\sim10$\,AU radius or even the infalling envelope in early stages, these curves mainly reflect the evolution of the entire protoplanetary disk and envelope from the largest positive $\alpha$ ($\sim5$ in our sample) to the photospheric value $\alpha\approx-3$.  

We can identify five $\alpha$ sub-regions that show distinct evolutionary characteristics in Fig.~\ref{fig:new_SED_classes} (we call it  \textit{SED staging scheme}): 
Stage~A $(2.5<\alpha)$ shows a very rapid decrease of $\alpha$ with time at the earliest stage so that very few sources are found in this stage (very short timescale of $\sim0.01$\,Myr) which should be true even after taking into account the sample incompleteness; 
Stage~B $(0<\alpha\le2.5)$ witnesses a gradual increase of star number towards the transition region from the Class~\textsc{i} to \textsc{f} (i.e., a gradual increase of disk dissipation timescale with time or a gradual slowdown of the decrease of $\alpha$) in a period of $\sim0.5$\,Myr; 
Stage~C $(-1<\alpha\le0)$ covers the ascending half of the $\alpha$-histogram peak which signals a rapid increase of star number toward the middle of the Class\,\textsc{ii} (i.e., an rapid increase of disk evolution timescale over time or a significant slowdown of the disk dissipation) in a period of $\sim1$\,Myr;
Stage~D  $(-2<\alpha\le-1)$ strides across the Class\,\textsc{ii} and \textsc{iii}, oppositely containing the descending half of the $\alpha$-histogram peak which indicates a sharp decline of star number (i.e., a sharp decrease of disk evolution timescale or a fast speedup of the dissipation of the remaining disk until its IR excess almost disappears) within another period of $\sim1$\,Myr;
Stage~E $(\alpha\le-2)$ brackets the diskless PMS stars and fainter debris disks (the timescale is difficult to estimate due to the lack of observed star samples and the potential confusion with MS stars, as discussed above. 

\begin{figure*}[ht!]
\epsscale{1.15} 
\plotone{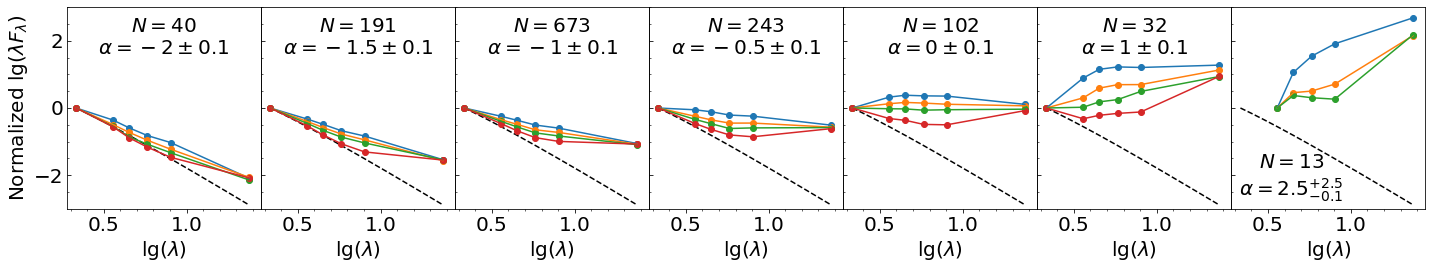}
\caption{Representative median SEDs normalized to 2MASS Ks band (or IRAC 3.6\,$\mu$m band when the Ks band is non-detection) around each border line of the SED stages in Fig.~\ref{fig:new_SED_classes} and inside Stages~B, C and D (see the selected $\alpha$ ranges and the number $N$ of YSOs in each group in the figure legends). The method to define the median SEDs can be seen in Sect.~\ref{subsubsec:SED_staging}. The sequence of panels are in the same order of $\alpha$ values as in Fig.~\ref{fig:new_SED_classes}.
Blue, yellow, green and red colors are used sequentially for the median SEDs from the most convex to the most concave in each panel.
The black dashed curve is a 6000\,K black body.
\label{fig:median_SED}}
\end{figure*}

The five SED Stages above very likely correspond to distinctive dissipation processes of the circumstellar dust at the level of YSO population. 
According to the current theory, a newly formed single star first appears as a Class\,0 protostar that has its youngest disk still embedded in a gas envelope \citep{Andre1993}. 
The envelope will be lost during Class\,\textsc{i} stage \citep{Lada1987}. 
The disk starts its long journey of dissipation mainly through the so-called UV-switch mechanism \citep{Clarke2001} in which the disk first loses mass through viscous accretion onto the star and then FUV, EUV and X-ray evaporation of the outer flaring disk will take over to switch off the mass supply to the inner accretion disk and leave behind an inner hole \citep[a transitional disk; e.g.,][]{Alexander2006,Gorti2009}. 
At the same time, dust grain settling towards the mid-plane of the disk \citep[e.g.,][]{Dullemond2004} and its growth to big grains and planetesimals \citep[e.g.,][]{DAlessio2001} should proceed in parallel.
The evolution of disks may have started quite early within the first 1\,Myr \citep{Furlan2009}. 
Thus, the earlier part of the disk dissipation should have started during Class\,\textsc{i} while its later part roughly corresponds to Class\,\textsc{ii} (classical T~Tauri stars; CTTS).
Finally, the photoevaporation, together with other processes such as stellar accretion, grain growth and planet formation, will quickly dissipate the remaining primordial disk during Class\,\textsc{iii} stage and a CTTS rapidly evolves through a weak-line TTS stage \citep[WTTS; see e.g.,][]{Alexander2006,Williams2011}. 
The observed dissipation timescale of the entire primordial disk is a little bit controversial in literature, ranging from $4\sim8$\,Myr \citep{Haisch2001,Ribas2014} down to $\sim1$\,Myr \citep{Cieza2007} and mostly around 2\,Myr \citep{Spezzi2008,Mamajek2009,Richert2018}. 
Further on, some planetary systems may develop a debris disk through collisions of planetesimals on longer timescales $\ge10$\,Myr \citep{Wyatt2008}. 

However, it has been suspected that not all disks follow the unique path of evolution delineated above. 
This was mainly inferred from the diversities of IR SEDs and millimeter maps \citep[e.g.,][]{Furlan2006,Cieza2007,Owen2016}. 
The diversity of SEDs is also found in our sample of YSOs. 
We present the median SEDs of several subgroups of YSOs near the border lines of the SED stages and inside Stages~B, C and D in Fig.~\ref{fig:median_SED}.
To create the median SEDs, we first fit a second order polynomial to each SED, then sort them according to the concavity (the second derivative) of the polynomial, and finally sequentially divide them into 3 or 4 roughly equal subgroups to compute median SEDs. 
Because YSOs around the border of Stages~A and B is few, we include all Stage~A YSOs into this group. 
The SEDs of this group have no detection in 2MASS Ks band so that they have to be normalized to the $3.6\mu$m band. 
The $\alpha=1\pm0.1$ group also has some Ks-band non-dections and only the 36 member stars with Ks-band flux are considered.
In addition, some member stars in the Orion region have no $24\mu$m flux due to the MIPS saturation issue discussed in Sect.~\ref{subsec:data_saturation}. 
They are omitted in the computation of the median SEDs, which mainly impacts Stage~C ($\alpha=-0.5\pm0.1$ and $-1\pm0.1$).
We will briefly discuss the major trends below, while more details will be analyzed in a dedicated paper.

As can be seen among the median SEDs, the diversity has already been large from the very beginning in Stage~A and begins to diminish from the end of Stage~B ($\alpha=0$) and on. 
A low level of SED diversity is maintained throughout Stage~D.
The well known transitional disks discovered since the work of \citet{Strom1989,Skrutskie1990} should have the most concave SEDs in our sample, because they usually show little excessive emission at NIR and MIR ($\lambda<20\,\mu$m) but maintain significant excess beyond $20\,\mu$m as optically thick full disks do \citep{Espaillat2014}. They mainly appear in our Stages~C and D and are all wrapped in the most concave median SED in the plots (the bottom red curves in the panels with $\alpha=-0.5\pm0.1$, $-1.0\pm0.1$, and $-1.5\pm0.1$). Thus, no more than $25\%$ of the sample YSOs in any $\alpha$ sub-range of the two stages belong to transitional disks.

The SED diversity during Stage~A can be attributed mainly to geometrical features of a standard `envelope+disk' model. According to the SED evolutionary models of \citet{Whitney2003}, the flaring disks in the earliest evolutionary stage should be embedded in an opaque infalling envelope with bipolar cavities. 
The opening angle of the cavities and the viewing angle to the cavity axis have strong impact to the observed SED morphologies. The model SEDs seen in different viewing angles in their {\rm late Class\,0 and Class\,\textsc{i}} cases (their Fig.~3) are sufficient to explain the median SEDs we found.

The SED diversity during Stage~B needs more factors to account for. 
The median SEDs show convex, flat and concave shapes. The latter two shapes can be similarly explained with the model SEDs viewed at different inclination angles in the {\rm Class\,\textsc{i} and late Class\,\textsc{i}} cases of \citet{Whitney2003}. 
However, the convex SEDs need either additional emission in the $3\sim8\,\mu$m range or deficit of emission in Ks and/or $24\mu$m bands to explain. 
This cannot be interpreted solely by the puff up of disk inner rim, because it mainly enhances emission around $2\sim3\,\mu$m for massive disks but not for T Tauri stars \citep{Dullemond2001}. 
We speculate the following candidate mechanisms for further work to test: 
1) these YSOs with convex SEDs are likely heavily embedded sources so that the uncorrected local extinction preferentially attenuates the light at shorter wavelengths; 
2) they could be binaries that have truncated the outer disk and reduced the $24\,\mu$m flux \citep{Williams2011}. 
However, only the local extinction scenario has the potential to also explain the diminishing trend of SED diversity towards later stages.

Because the SED diversities in the early stages can be satisfactorily interpreted as above by different cavity opening angles and viewing angles, the issue of multiple disk evolutionary paths mainly exists in the later evolutionary stages. 
\citet{Wood2002,Lada2006} proposed a `homologously depleted disk' or `anemic disk' model for a large population of observed {\rm Class\,\textsc{ii}} disk sources that always maintain a power law shape of SED during the evolution. 
Our Fig.~\ref{fig:new_SED_classes} also shows that the largest portion of the {\rm Class\,\textsc{ii} and \textsc{iii} YSOs in GB} do have power-law SED shapes. 
Such disks cannot be interpreted alone by grain settling and growth and stellar accretion, because the fast dust settling and coagulation must be balanced by replenishment through turbulence in the disk \citep{Dullemond2005} and the accretion may be too slow in later stages to explain the observed fast disk dispersal. 
However, combining them with the self-shadowing effects of an inner wall \citep[it reduces the far IR disk emission;][]{Dullemond2004} and photoevaporation \citep{Alexander2006} has the potential to explain the entire observed evolution sequence of anemic disks.
On the other hand, a small number of millimeter-strong transition disks \citep{Owen2016} represent another distinct evolution pathway of some {\rm Class\,\textsc{ii} and \textsc{iii} objects} which may need the formation of giant planets to clear out the inner disk within $2\sim3$\,Myr \citep[e.g.,][]{Rice2003,Varniere2006,Dodson-Robinson2011,Andrews2011,Gorti2011,Dong2015,Sallum2015}. 
The existence of multiple pathways may pose a question on the short lifetime of transitional disks that was constrained by counting stars also in the `anemic disk' channel.
The stars' choice between the two evolutionary paths might depend on the details of planet formation, which we will discuss in the next subsection.

Referring to the above picture, our observed SED evolutionary stages can be matched to disk dissipation processes as follows. 
The very few sources observed in Stage~A are likely stars with infalling envelope actively interacting with bipolar jets.
This is a brief stage of $\sim10^4$\,yrs for the newly born stars to emerge at NIR. 
The viewing angle of the bipolar cavities strongly impact the detectability of the stars and their SED diversity.

Detectable disk sources begins to build up during Stage~B. 
In this $\sim0.5$\,Myr process, it is natural to expect the dissipation of the remaining envelope by widening the bipolar cavities, grain settling toward the mid-plane and grain growth. The flattening of the disk results in lower surface temperature, which explains the observed quick decrease of $\alpha$.
The disk flattening slows down when the disk becomes flatter, giving rise to the observed gentle rise up of the $\alpha$ histogram with decreasing $\alpha$.
The SED diversity due to the different bipolar cavity opening angles and viewing angles also begins to diminish with disk flattening.
We suggest that the creation of an inner-disk wall might have started at a certain moment because the strong stellar accretion should be powerful enough to generate energetic radiation to blow up the wall simultaneously. 

After then, the SED evolution further slows down significantly in Stage~C. 
This is because a stable flaring disk structure is established during this $\sim1$\,Myr stage. 
The disk should have an inner wall that shadows and protects a large part of the flaring disk. 
The disk evolution is mainly driven by the slow stellar accretion through the wall. 
Dust grain settling and growth is balanced by turbulent grain replenishment. 
Planetesimals and planets begin to form in the mid-plane. 
This stage ends at a yet poorly understood critical point around the $\alpha=-1$ peak in which the slowing trend of disk dissipation reverses to an accelerating trend (we will discuss the mechanism in next subsection).
The sub-millimeter mapping of YSOs in the Taurus-Auriga SF region by \citet{Andrews2005} has uncovered a sharp decrease of disk mass during the Class\,\textsc{ii} stage (their Fig.~15), which also supports the $\alpha$ peak to be the critical starting point of the final runaway disk gas dissipation.

Stage~D should be the final stage of the primordial disk. The further dissipation of the entire disk speeds up, possibly due to the (partial) clearing of the inner disk. 
In the average $\alpha$ histogram, the profile shape of this stage is roughly left-right symmetrical to that of Stage~C, indicating a nearly reciprocal timing of the two stages.
The division of Stages~C and D is thus very different from the so-called two-time-scale division between the long full disk stage {\rm (roughly Class\,\textsc{ii})} and the brief transitional disk stage  {\rm \citep[roughly Class\,\textsc{iii}; e.g.,][]{Alexander2014,Ercolano2017}.} Therefore, a different physical interpretation is needed here. 
We propose that, due to some unknown processes at the $\alpha=-1$ peak,
the inner wall begins to collapse and the disk is quickly dispersed by the combination of stellar accretion and photoevaporation in the $\sim1$\,Myr period of this stage. 
The other processes such as grain settling and growth only play minor roles due to the diminishing gas and dust density.

Finally, the YSOs become diskless stars or debris-disks in Stage~E.
We comment that some debris disks could return from Stage~E to Stage~D if their $24\mu$m dust emission would be strong enough. However, identifying such objects in our sample is beyond the scope of this work.

\subsubsection{Role of giant planet formation in disk dispersal} \label{subsubsec:planets}

The paramount $\alpha=-1$ histogram peak (in the middle of the traditional Class\,\textsc{ii}) is not only the slowest disk dissipation moment, but the critical point when the disk dispersal reverses from slowing down to speeding up.
No mechanism is known for this reversal.
Processes such as stellar accretion, dust settling and grain growth are not expected to show such a critical state. 
Photoevaporation does possess a critical point when the evaporation mass loss rate overtakes the accretion rate and an inner hole will be cleared out and the outer disk will be shed quickly. 
However, this critical point is expected only after several Myr of stellar accretion \citep[e.g.,][]{Clarke2001,Alexander2006,Gorti2009}, which is too long compared to the age of the $\alpha=-1$ peak (1.76\,Myr). In addition, this is the mechanism for the two-time-scale problem of disk dispersal, as mentioned in Sect.~\ref{subsubsec:SED_staging}, which does not agree to the observed symmetry in the average SED histogram between Stages~C and D.


The only hopeful driving process is the widely studied giant planet formation. 
The giant planets in the Solar system, Jupiter and Saturn, have a huge gas sphere so that they must accrete enough gas before the dissipation of the gas disk.
Therefore, in our average $\alpha$ histogram, the giant planets of low mass stars very likely mature around the $\alpha=-1$ peak.
\citet{Ercolano2017} reached a similar conclusion in their review of disk dispersal and planet formation and even suggested to call {\rm Class\,\textsc{ii}} sources as planet forming disks, instead of protoplanetary disks.
The capability of giant planets in clearing the inner disk (so as to speed up the dispersal of the outer disk) is also supported by observations. For example, \citet{Andrews2011} found that only dynamical clearing by giant planets (or brown dwarfs) can explain the cavities in 12 transition disks they mapped at sub-millimeter; 
\citet{Sallum2015} caught accreting planets in action in the multiple ring system of the transition disk source LkCa~15.
Although the formation of giant planets alone cannot explain the disappearance of disks because of their too low masses \citep{Pascucci2010}, it can play a key role in combination with photoevaporation \citep{Rosotti2013}. 

In order to better understand the capability of giant planets in speeding up disk dispersal, let us recall the typical three step scenario of giant planet formation \citep{Pollack1996}.
In the first step, a runaway accretion of cm-sized pebbles builds up an initial solid core above $\sim10\,{\rm M}_\oplus$ to enable gas accretion \citep{Lambrechts2012}. 
The second step is the most time consuming: the gas accretion will quickly clear up the feeding zone of the protoplanet to greatly slow down further accretion of solids and the stable gas sphere cannot grow further by adding more gas.
However, the accreting solid core will dynamically migrate radially to maintain a non-negligible accretion rate; 
it will achieve a critical cross-over mass threshold (i.e., the accreted planetesimal mass equals the accreted gas mass) within about $1\sim3$\,Myr \citep{Alibert2005,Cieza2012,Bitsch2015}. Finally, a runaway gas accretion is triggered \citep[because the accreted gas sphere starts to gravitationally collapse;][]{Mizuno1978} to complete the giant planet formation rapidly. 

The runaway gas accretion in the last step has the capability to open a gap in both the dust and gas disks \citep{Crida2006,Fung2014} and to play a key role in reversing the slowing trend of the disk dissipation. 
Although the planet opened gaps may not totally block out the supply of  gas and small dust grains to the inner accretion disk \citep[e.g.,][]{Rice2006,Crida2007,Zhu2012}, it can significantly reduce the gas density of the inner disk to speed up the dispersal of the outer disk by photoevaporation \citep{Dodson-Robinson2011,Andrews2011,Rosotti2013}.
This speed-up commences around $\alpha=-1$ according to our average $\alpha$ histogram.

However, the above giant planet formation scenario is only obviously applicable to transitional disks, but not to anemic disks that do not show the signature of inner disk hole in their SED morphologies.
There is no readily available model to explain why these disks also speed up their dispersal around $\alpha=-1$. 
We speculate that the giant planet formation might still work in this case, but the planet masses could be lower than in the transitional disks so that the shallower gaps and inner holes are not discernible in SEDs. 
However, the shallow gaps and holes still reduce the mass supply to the inner disk to speed up the photoevaporation. 
Of course, other processes such as grain settling and growth could be contributing too.

There have been some evidences supporting the planet scenario for anemic disks.
On the one hand, \citet{Najita2007,Sicilia-Aguilar2008,Muzerolle2010} found that anemic disks tend to appear around lower mass stars with lower disk masses while transitional disks around more massive stars with higher disk masses, indicating that the former tend to form less massive giant planets than the latter.
This is also consistent with the very low occurrence rates of massive giant planets in direct imaging observations \citep[e.g.,][]{Bowler2016}. 
On the other hand, imaging studies have been providing acuter clues. 
The ALMA large program DSHARP \citep{Andrews2018} have revealed that 16 out of the 20 nearby Class\,\textsc{ii} anemic disk targets show gaps and/or spirals related to planet dynamics.
The ALMA archival study of \citet{Francis2020} have also shown that many disks with gaps and inner holes at millimeter wavelengths have their SED shapes actually closer to anemic disks than to typical transitional disks (their Fig.~16). 
With more ALMA continuum or CO line maps, \citet{Long2018,Wolfer2022} further proposed that many of the disk sources (including transitional disks) have their disk gaps possibly opened by low mass planets. 
All of them hint on the ubiquitous role of planet formation also in anemic disks defined from SEDs.

If the ubiquitous role of planet formation in anemic disks would be further confirmed, there emerges a simple uniform picture in which only the rare most massive giant planets (or even stellar companions) can clear out the massive inner dust disks to leave behind the small number of millimeter bright transitional disks reviewed by \citet{Owen2016}, while the more common less massive planets grow in and help disperse the larger population of anemic disks.

Finally, we comment that the current theory of giant planet formation still has the room for improvements. For example, many millimeter-bright transitional disks with a large inner disk hole are found to show sizable gas accretion \citep{Manara2014,Owen2016}, whilst giant planet formation and photoevaporation models constantly predict too many non-accreting transition disks \citep{Rosotti2015,Ercolano2021}.
Further studies are desired to solve such conundrums and to explain the $\alpha=-1$ peak identified here.

\section{Summary} \label{sec:summary}

By reprocessing in a uniform manner the IR SEDs of all member stars of Gould's Belt (GB) protoclusters surveyed by Spitzer Space Telescope, we have successfully identified the common features in the histograms of SED slopes ($\alpha$) that represent the distribution of disk evolution life time over the $\alpha$ value. This has allowed us to convert $\alpha$ into disk evolution age to obtain the first panoramic view to the large fluctuations of SFRs in the GB protoclusters in the past 3\,Myr history. We have found the following properties:\\
(1) The star formation is roughly continuous in GB and the amplitudes of SFR fluctuation are $20\%\sim60\%$, or $\sim40\%$ on average;\\
(2) Spatially close protoclusters tend to share similar SFR fluctuation trends, indicating common driving forces for the fluctuations at large size scales;\\
(3) The amplitudes of fluctuation are likely independent of the average SFR and the number of SF episodes.

The average $\alpha$ histogram is largely free of sample selection bias. 
It shows several statistic features that allow us to define five SED evolutionary stages: A, B, C, D and E, with their dividing points at $\alpha=2.5$, 0, -1 and -2 or at negative SED evolution ages of -0.01, -0.55, -1.76 and -2.84\,Myr. It can be considered as a further development of the traditional SED classification scheme of {\rm \textsc{0,i,f,ii} and \textsc{iii}} that was based on different $\alpha$ values and on the theoretical meaning of the different SED classes. 

The SEDs of individual YSOs show diversities. 
The diversities in the earlier Stages~A, B and C can be interpreted by different opening angles and viewing angles of bipolar cavities, while the diversities during Stage~D indicates two disk evolution pathways: anemic and transitional disks. 
Despite the diversities, the SED staging scheme provides us a convenient time frame of disk evolution to coordinate further studies.
The SED staging scheme can be better matched to the dissipation stages of circumstellar matter as follows: \\
(1) bipolar cavities begin to develop in the infalling envelope in the first $10^4$\,yr; \\
(2) the infalling envelope is dissipated by stellar jets and winds and grain settling and growth, and possibly an inner wall is puffed up within $\sim0.5$\,Myr; \\
(3) a stable flaring disk structure protected by the shadowing of an inner wall is developed and stellar accretion in the inner disk, photoevaporation of the inner wall, grain settling, growth and planet formation in the disk mid-plane jointly drive its slow evolution in $\sim1$\,Myr; \\
(4) giant planets form and open gaps in the disk to starve the inner accretion disk, which triggers the acceleration of the final disk dispersal by photoevaporation in another $\sim1$\,Myr; \\
(5) the final stage of diskless star or debris disk is not well constrained in this work.\\

This work has also opened up the opportunity to have a better look into the evolution of the cluster spatial structures with time. We will report it in the next paper.

\begin{acknowledgments}
We thank the science referee for helping us improving several important defects and pushing us to take into account the diversities of SED and disk evolution paths, which has significantly improved our data analysis and discussions. 

We thank Prof. Diego Mardones for many useful discussions in the early stage of the work. ML jointly initiated the project, performed all data collection, processing, analysis, code writing and participated in the manuscript revision.
JH jointly initiated the project, guided every step of the entire work, wrote and revised the manuscript.
The other co-authors participated many discussions and significantly improved many aspects of the data analysis and science discussions.

JH and TL acknowledge the support from the National Natural Science Foundation of China (NSFC) through grant Nos. 11873086, 12073061 and 12122307. TL also thanks the International Partnership Program of the Chinese Academy of Sciences (CAS) through grant No. 114231KYSB20200009, the Shanghai Pujiang Program (20PJ1415500), and science research grants from the China Manned Space Project with no. CMS-CSST-2021-B06. XL is supported by China Postdoctoral Science Foundation project (2021M703099). This work is sponsored (in part) by the Chinese Academy of Sciences (CAS), through a grant to the CAS South America Center for Astronomy (CASSACA) in Santiago, Chile.

This research made use of the cross-match service provided by CDS, Strasbourg.

\end{acknowledgments}

\vspace{5mm}
\facilities{Spitzer \citep[NASA;][]{Werner2004}, GAIA \citep[ESA;][]{Gaia2016}, 2MASS \citep{Skrutskie2006}.}

\appendix

\section{Recovery of saturated MIPS fluxes in Orion region}\label{sec:24umflux}

The \textit{Spitzer}/MIPS maps of the Orion regions suffer from saturation issues in a few brightest areas so that no reliable 24\,$\mu$m (in some cases also no IRAC 5.8 and 8.0$\mu$m) photometry is available for a significant fraction of member stars in those protoclusters \citep[Orion\,A, B-S and B-M;][]{Megeath2012}. To partially remedy the bias resulted from the missing of these member stars, we need an approach to reconstruct their $\alpha$($2.2\sim24\,\mu{\rm m}$) values from available data at shorter IR wavelengths. In each of the three protoclusters, we first compare the histograms of $\alpha$ values defined in $2.2\sim5.8$ and $2.2\sim8.0\,\mu$m wavelength ranges between the member stars inside and outside of the saturated areas and find that they are statistically similar. It is thus reasonable to assume that the relationship of the $\alpha$ values defined in the three wavelength ranges are statistically similar for all member stars, no matter whether they suffer from the MIPS saturation issue. Then, for those member stars outside of the saturated areas, we compare their $\alpha$ values defined in different wavelength ranges and find the following good linear correlations in all cases. 
\begin{itemize}
    \item For Orion~A: \\
    \begin{equation}
        \alpha(2.2\sim24\,\mu{\rm m})=0.48(\pm0.01)\times\alpha(2.2\sim5.8\,\mu{\rm m})-0.13(\pm0.02),
    \end{equation}
    \begin{equation}
        \alpha(2.2\sim24\,\mu{\rm m})=0.59(\pm0.01)\times\alpha(2.2\sim8.0\,\mu{\rm m})-0.07(\pm0.01).
    \end{equation}
    \item For Orion~B-M: \\
    \begin{equation}
        \alpha(2.2\sim24\,\mu{\rm m})=0.55(\pm0.02)\times\alpha(2.2\sim5.8\,\mu{\rm m})-0.04(\pm0.04),
    \end{equation}
    \begin{equation}
        \alpha(2.2\sim24\,\mu{\rm m})=0.66(\pm0.03)\times\alpha(2.2\sim8.0\,\mu{\rm m})+0.01(\pm0.04).
    \end{equation}
    \item For Orion~B-S: \\
    \begin{equation}
        \alpha(2.2\sim24\,\mu{\rm m})=0.51(\pm0.05)\times\alpha(2.2\sim5.8\,\mu{\rm m})-0.13(\pm0.08),
    \end{equation}
    \begin{equation}
        \alpha(2.2\sim24\,\mu{\rm m})=0.61(\pm0.05)\times\alpha(2.2\sim8.0\,\mu{\rm m})-0.09(\pm0.06).
    \end{equation}
    
\end{itemize}
They allow us to infer the needed $\alpha$($2.2\sim24\,\mu{\rm m}$) values of the saturated member stars from their available $\alpha$($2.2\sim8.0\,\mu{\rm m}$) or $\alpha$($2.2\sim5.8\,\mu{\rm m}$) values. As a result, the $\alpha$($2.2\sim24\,\mu{\rm m}$) values are recovered from $\alpha$($2.2\sim8.0\,\mu{\rm m}$) for 502 (19\%), 36 (13\%) and 78 (24\%) member stars that have available 8.0$\mu$m flux and for another 294 (11\%), 21 (8\%) and 62 (19\%) member stars that have both the 8.0 and 24\,$\mu$ fluxes missing but have the 5.8\,$\mu$m flux available for the three protoclusters respectively (the percentages in the parentheses are relative to the total number of member stars). Nevertheless, we still have to omit the rest 257 (10\%), 8 (3\%) and 61 (19\%) member stars that have all the 5.8, 8.0 and 24\,$\mu$m fluxes missing in the three protoclusters respectively. 

\section{Error bars of the SED-slope-to-disk-age conversion curve}
\label{app:alpha_time_error_bar}
The error bars of the SED slope ($\alpha$)-disk age conversion curve in Fig.~\ref{fig:alpha_to_age} cannot be analytically obtained from the error bars of the average $\alpha$-histogram, because it involves integration of the latter. We obtain them by numerically re-sampling the average $\alpha$-histogram to imitate its error bars. 

First, we compute the correlation matrix of the 14 $\alpha$-bins of the average $\alpha$-histogram from the original $\alpha$-histograms of the 13 protoclusters. This step is necessary because the different $\alpha$-bins are not necessarily independent when being averaged over the 13 protoclusters. The resulting correlation coefficients are shown in Appendix Table~\ref{tab:corr_matrix}. The only valuable comment we can make is that the bins 5, 6, 7, 8, 9, 10 and 12 ($\alpha=0.25\sim3.75$) are significantly positively correlated. This could be explained by the short disk life times in these bins so that these bins can be under the control of a single SF episode. This fact confirms our preliminary finding of the positive correlation of star number counting between Class\,\textsc{f} and \textsc{0+i} protostars (not published).

Then, we generate $10^4$ random realizations of the average $\alpha$-histogram using this observationally constrained correlation matrix (actually the covariance matrix is used) in a multi-variate Gaussian distribution. Each random histogram is similarly integrated as in Eq.~\ref{eq:alpha-t} to obtain a random copy of the $\alpha$-age conversion curve. The dispersion of the $10^4$ random samples of the conversion curve just gives the error bar at any given $\alpha$ value in Fig.~\ref{fig:alpha_to_age}. The computed results are given in Appendix Table~\ref{tab:alpha_time}.

\section{Appendix tables used in the paper or behind the plots}
\label{app:tables}
\begin{table}[!htb]
\begin{center}
\begin{minipage}{190pt}
\centering
\caption{Distance range used as criterion for member stars of each Gould's Belt protocluster and the numbers of removed foreground ($n_{\rm fg}$) and background ($n_{\rm bg}$) contaminating stars.}\label{tab:dist_range}
\begin{tabular}{@{}llrr@{}}
\toprule
Protocluster & Distance range (pc) & $n_{\rm fg}$ & $n_{\rm bg}$ \\
\midrule
Serpens NE   & $400\sim600$ & 0 & 102\\
Aquila-Main  & $400\sim600$ & 4 & 350\\
Ophiuchus    & $100\sim200$ & 0 & 14\\
Serpens Main & $300\sim600$ & 2 & 24\\
Chamaeleon I & $150\sim300$ & 0 & 3\\
Perseus-E    & $200\sim400$ & 1 & 10\\
Taurus       & $100\sim200$ & 2 & 84\\
IC5146-E     & $600\sim900$ & 0 & 17\\
Orion A      & $300\sim600$ & 22 & 98\\
Orion B-S    & $300\sim600$ & 3 & 8\\
Orion B-M    & $300\sim600$ & 2 & 16\\
AurigaCMC    & $400\sim600$ & 2 & 11\\
Perseus-W    & $200\sim400$ & 2 & 7\\
\botrule
\end{tabular}
\end{minipage}
\end{center}
\end{table}

\begin{table}[htb]
\begin{center}
\begin{minipage}{250pt}
\centering
\caption{Data befind the plots: Columns 1-3 are for the average $\alpha$ histogram （after spline interpolation and normalized to an area of unity) and error bars ($\alpha$, $p$ and $\sigma_p$) in the left half of Fig.~\ref{fig:alpha_histo}; columns 4-6 are for the $\alpha$-age conversion curve and error bars ($\alpha$, age and $\sigma_{\rm age}$) in Fig.~\ref{fig:alpha_to_age}. The former curve is sampled at the bin centers of the histogram while the latter is at the borders.}\label{tab:alpha_time}
\begin{tabular}{@{}rll|rll@{}}
\toprule
$\alpha$ & P & $\sigma_P$ &  $\alpha$ & age(Myr) & $\sigma_{\rm age}$(Myr)\\
\midrule
-2.25 & 0.0E+00$^a$ & 0.0E+00 & -2.0 & -2.8E+00 & 3.9E-01 \\
-1.75 & 1.4E-01 & 1.1E-01 & -1.5 & -2.6E+00 & 3.7E-01 \\
-1.25 & 6.3E-01 & 1.6E-01 & -1.0 & -1.8E+00 & 4.4E-01 \\
-0.75 & 6.2E-01 & 1.9E-01 & -0.5 & -9.0E-01 & 3.8E-01 \\
-0.25 & 2.2E-01 & 5.9E-02 & 0.0 & -5.5E-01 & 3.3E-01 \\
0.25 & 1.5E-01 & 4.5E-02 & 0.5 & -3.3E-01 & 2.5E-01 \\
0.75 & 1.0E-01 & 5.1E-02 & 1.0 & -1.9E-01 & 1.9E-01 \\
1.25 & 6.6E-02 & 4.6E-02 & 1.5 & -9.2E-02 & 1.3E-01 \\
1.75 & 4.2E-02 & 3.6E-02 & 2.0 & -3.3E-02 & 7.1E-02 \\
2.25 & 1.6E-02 & 2.9E-02 & 2.5 & -1.0E-02 & 2.7E-02 \\
2.75 & 4.5E-03 & 9.8E-03 & 3.0 & -3.2E-03 & 1.0E-02 \\
3.25 & 1.0E-03 & 3.2E-03 & 3.5 & -1.5E-03 & 4.8E-03 \\
3.75 & 9.3E-04 & 3.2E-03 & 4.0 & -3.3E-04 & 1.2E-03 \\
4.25 & 0.0E+00 & 0.0E+00 & 4.5 & -2.1E-04 & 1.1E-03 \\
4.75 & 6.2E-04$^b$ & 2.1E-03 & 5.0 & 0.0E+00 & 0.0E+00 \\\botrule
\end{tabular}\\
$^a$The $\alpha=-2.25$ histogram bin is an artificial one added to avoid the histogram from going to zero at $\alpha=-2$.\\
$^b$The small positive histogram value of 6.2E-04 is due to the oscillation of the spline fitting function, which has little impact to this work.
\end{minipage}
\end{center}
\end{table}

\begin{table}[]
\begin{center}
\begin{minipage}{420pt}
\centering
\caption{Correlation coefficients among the 14 bins of the average $\alpha$ histogram in Appendix~\ref{app:alpha_time_error_bar} and Appendix Table~\ref{tab:alpha_time} (excluding the artificial bin $\alpha=-2.25$). Bins 1 and 14 correspond to $\alpha=-1.75$ and 4.75, respectively. The correlation coefficients with absolute values $\ge0.5$ are highlighted in bold face.\label{tab:corr_matrix}}
\begin{tabular}{r|rrrrrrrrrrrrrr}
1   & 1.0  \\
2   & 0.43          & 1.0 \\
3   & {\bf-0.57}   & {\bf-0.56} & 1.0 \\
4   & {\bf-0.5}    & -0.09      & 0.084     & 1.0 \\
5   & 0.3           & -0.37     & -0.49     & -0.3      & 1.0 \\
6   & -0.39     & {\bf-0.5}     & 0.00079   & 0.19      & 0.39      & 1.0 \\
7   & -0.36     & -0.4          & -0.14     & 0.22      & 0.48      & 0.31  & 1.0 \\
8   & -0.21     & {\bf-0.53}    & -0.19     & -0.16     & {\bf0.71} & 0.43  & {\bf0.56}     & 1.0 \\
9   & -0.3      & -0.47         & -0.14     & -0.12     & {\bf0.6}  & 0.36  & 0.47          & {\bf0.9}  & 1.0 \\
10  & -0.22     & -0.27         & -0.22     & -0.09     & 0.47      & 0.27  & 0.18          & {\bf0.82} & {\bf0.91} & 1.0 \\
11  & -0.22     & -0.44         & -0.11     & -0.14     & {\bf0.56} & 0.38  & 0.18          & {\bf0.82} & {\bf0.94} & {\bf0.94} & 1.0 \\
12  & -0.2      & -0.43         & -0.13     & -0.15     & {\bf0.57} & 0.39  & 0.19          & {\bf0.82} & {\bf0.94} & {\bf0.94} & {\bf1.0}      & 1.0 \\
13  & 0.0       & 0.0           & 0.0       & 0.0       & 0.0       & 0.0   & 0.0           & 0.0       & 0.0       & 0.0       & 0.0           & 0.0       & 1.0 \\
14  & -0.026    & {\bf0.52}     & -0.34     & {\bf0.52} & -0.37     & -0.17 & -0.21         & -0.27     & -0.16     & 0.1       &-0.089         & -0.083    & 0.0   & 1.0 \\\midrule
Bin & 1          & 2          & 3     & 4         & 5         & 6     & 7         & 8         & 9         & 10        & 11    & 12    & 13 & 14 \\
\end{tabular}
\end{minipage}
\end{center}
\end{table}

\bibliography{GB_SFR_history}{}
\bibliographystyle{aasjournal}

\end{CJK*}
\end{document}